# Science and Philosophy: A Love-Hate Relationship

Sebastian de Haro

Trinity College, Cambridge, CB2 1TQ, United Kingdom
Department of History and Philosophy of Science
Free School Lane, Cambridge CB2 3RH, United Kingdom
sd696@cam.ac.uk

**Abstract**

In this paper I review the problematic relationship between science and philosophy; in particular, I will address the question of whether science needs philosophy, and I will offer some positive perspectives that should be helpful in developing a synergetic relationship between the two. I will review three lines of reasoning often employed in arguing that philosophy is useless for science: a) philosophy's death diagnosis ('philosophy is dead'); b) the historic-agnostic argument/challenge "show me examples where philosophy has been useful for science, for I don't know of any"; c) the division of property argument (or: philosophy and science have different subject matters, therefore philosophy is useless for science).

These arguments will be countered with three contentions to the effect that the natural sciences need philosophy. I will: a) point to the fallacy of anti-philosophicalism (or: 'in order to deny the need for philosophy, one must do philosophy') and examine the role of paradigms and presuppositions (or: why science can't live without philosophy); b) point out why the historical argument fails (in an example from quantum mechanics, alive and kicking); c) briefly sketch some domains of intersection of science and philosophy and how the two can have mutual synergy. I will conclude with some implications of this synergetic relationship between science and philosophy for the liberal arts and sciences.

## 1. Introduction

In this paper I will argue that: (i) The natural sciences need philosophy; and (ii) That scient*ists* need philosophy. I will also address some possible consequences of these theses for the Liberal Arts and Sciences. In doing so, I will have to define the sense in which I mean that science 'needs' philosophy and make a distinction between different ways in which different aspects or branches of science need philosophy. Most of my examples will be from physics. This being part of my professional bias, I claim that the arguments that apply to physics apply to biology, earth science, and other natural sciences as well. As I will argue, the most important distinction to be made is not between one natural science and another, but between *fundamental* and *applied* science. Once this distinction is made, the harm of treating all of the sciences *en bloc,* on the model of physics, can be minimized.

Why should I be defending the use of philosophy? After all, the thesis that philosophy is useful for science is not likely to be agreed upon by all practicing scientists. Science, not philosophy, is widely regarded as the more secure source of knowledge. For it has a method for declaring theories wrong: in other words, for falsifying its results. This method is called Experiment. And science has given us machines, abundant energy, technology, and a healthy attitude of scepticism. The scientific worldview has freed us from prejudice, ignorance, and the ironclad rule of authority. Natural scientists, not philosophers, have earned the trust of the public opinion in matters of truth, learning, and understanding. Experimental results, and not the scholastic distinctions of the philosophers, are the final judges in the court of Science. This, at least, would seem to be a widely held view, and partly for good reasons. So why care about philosophy after all?

The relation between science and philosophy is an intricate and somewhat problematic subject,[1] as I will review in the next Section. On the one hand, some great scientists have been great philosophers—not necessarily in the professional sense, but in the sense of deep thinking: science and philosophy often went together in the work of great figures such as Newton and Leibniz, so that it is sometimes hard—and perhaps unnecessary, and certainly anachronistic—to say where science ends and where philosophy begins. But on the other hand, philosophy is often regarded as useless, so that a philosophical outlook is irrelevant for science at best, and harmful at worst—as evinced by long pages of armchair philosophy that is blissfully uninformed by science. Or so the story goes. Hence my topic of the 'love-hate relationship'.

Here I will concentrate on basic aspects of the relation between the two, and reduce the applications to education to a few final considerations. Reaching clarity about the fact that philosophy is useful for science is by itself an important and urgent task. Understanding this relationship is a first key step toward developing a synergetic relationship between the two fields.

## 2. Science doesn't need philosophy

Let us start with the objections to the first thesis above, that the sciences need philosophy. There are various good reasons why scientists can claim—and have claimed—that science does not need philosophy or that, more or less equivalently, philosophy is useless for science. I will consider three lines of reasoning here: the argument from the decline or death of philosophy, the historical or empirical argument, and the argument based on the contention that science and philosophy have different objects and methods.

### a) The death of philosophy

Stephen Hawking has declared the official 'death' of philosophy, seemingly echoing Nietzsche's famous 'God is dead'. Commenting on questions such as the behaviour of the universe and the nature of reality, Hawking writes: "Traditionally these are questions for philosophy, but philosophy is dead. Philosophy has not kept up with modern developments in science, particularly physics. Scientists have become the bearers of the torch of discovery in our quest for knowledge." (Hawking 2010, p. 5). In this argument, knowledge must be grounded on natural science. Questions such as, "what is the nature of physical reality", "what things are out there in the world?" are questions that used to be within the domain of philosophy, but are now part of science. Something in philosophy must therefore be missing, without which philosophy is left a 'dead' discipline. And when a discipline is dead, one might just as well ignore it.[2] Since philosophers haven't kept up with modern science, they have cut themselves off from our most secure source of knowledge and discovery. Hence the question: can we dismiss Hawking's provocative suggestion that philosophers have largely been neglecting the natural sciences, thereby manoeuvring themselves towards a margin of irrelevance, if in our world the natural sciences are becoming increasingly dominant?

### b) The historic-agnostic or empirical argument

The historic-agnostic argument is more cautious, and can be summarized as an agnostic stance about the usefulness of philosophy for science. It amounts to something like this: "I have never seen any examples of the usefulness of philosophy for science or, when I have seen usefulness in anything that philosophers were saying about science, it was because they were doing science not philosophy". The argument can be appended with an enumeration of instances where the limited scope of a philosophical framework hampered progress in science and perhaps also a theoretical account of why that was the case.

[1] For a different perspective on this topic, see Kitchener (1988).
[2] This is not what Hawking does, and the reason for it will become clear in the next section. He does not ignore philosophy, but engages with it.

Examples to this effect would seem to abound. Think of Plato's requirement, expressed in the *Timaeus,* that the movements of the planets should be taken to be based on uniform circular motions. This mathematical postulate was grounded on the philosophical and theological doctrine that the most perfect motion was circular, because the motion of the mind when it reflects upon itself is circular (Plato (T) 34a, 36c, 40a). It became apparent very early on that this axiom was untenable for concentric spheres. Hipparchus and Ptolemy had to add a contrived system of eccentrics and epicycles to save the phenomena. Nevertheless, they still formally adhered to the Platonic postulate, which has been seen by many as a hampering factor in the progress of cosmology (Dijksterhuis 1950, Part I, II D 15 and III C 68).

Another example could be Descartes' requirement that all of physics should be based on the mechanical interactions between corpuscula with no other properties than form, size, and quantity. The dictum that physical interactions ought to take place by local contact collided with Newton's theory of gravity, which entailed action at a distance. The dictum led Descartes to formulate his clever and imaginative, but arbitrary and unexplanatory theory of gravity on the basis of vortices, and a theory of magnetism based on the supposed screw-shapes of particles. Making the observed macroscopic phenomena supervene on microscopic details that were unobservable, and could therefore be amended at zero risk, allowed him to give qualitative and imaginative explanations of those phenomena: but he fell short of finding quantitative descriptions—let alone predictions. It took Newton to show, in Book II of the *Principia,* that Descartes' vortex theory was not only physically inconsistent—additional external forces would be required to keep the vortices moving—but also inconsistent with Kepler's laws. Richard Westfall gave the following fulminating evaluation of Descartes's philosophy in connection to mechanics: "Most of the major steps forward in mechanics during the [17th] century involved the contradiction of Descartes. Although the mechanical philosophy asserted that the particles of matter of which the universe is composed are governed in their motions by the laws of mechanics, the precise description of motions led repeatedly to conflict between the science of mechanics and the mechanical philosophy." (Westfall 1971, p. 138). Again, one might take this as an instance where philosophy constrains scientific progress by its adherence to pre-conceived and non-negotiable ontological ideas.

In a third, more recent example, Lawrence Krauss has argued that, when it comes to the most philosophical questions about for instance quantum mechanics, such as 'what is a measurement?', he finds the reflections of physicists more useful than those of philosophers (Krauss 2012), again reflecting the agnostic stance that says: "Show me examples where philosophy has been useful for science, for I don't know of any."

The historical argument, then, generally amounts to the following: "Look at the relationship between science and philosophy in the past. Any attempts at close collaboration or integration between science and philosophy have always failed. It is useless to try."

**c) Division of property: method and subject matters**

The third argument lies at the root of the other two. It says that philosophy and the natural sciences have different subject matters, therefore a small basis of overlap: they can live happily together without interfering with each other. This would explain the tendency of philosophers, signalled by Hawking, to retreat into the study of human affairs and human societies, leaving the study of nature to natural scientists.

The underlying reasoning can be understood as follows. The traditional distinction, at any rate since the 19$^{th}$ century, between the natural sciences and the humanities, is in their subject matters: nature would be the subject matter of the natural sciences, whereas humanities would busy themselves with the products of the human mind. The social sciences would then focus on human behaviour and social realities. Science would only be interested in brute matters of fact and not in social or linguistic constructs, and it would know those matters of fact by means of experiments carried out under certain conditions and subject to requirements of transparency and replicability. Placing philosophy in the camp of the humanities and the social sciences as opposed to the natural sciences: it also would deal with products of the mind and social constructs. This would both institutionalise philosophy's independence from science, as well as establish its uselessness for science.

To this difference in subject matters corresponds a difference in method, emphasized by Wilhelm Dilthey: *erklären* (to explain) would be the task of the natural sciences; the humanities would instead aim at *verstehen* (to understand, or comprehend): not to give a reductive account in terms of causal relation, but to create a comprehensive view where parts can be related to the whole. Science would aim at formulating general laws of nature via the universal language of mathematics; a universal language that, even if it would include probabilistic laws, would admit of no ambiguities; the goal of the scientist would then be: to explain the behaviour of nature in terms of laws that can be falsified or verified by experiments. It should be said that this methodological argument can be held quite independently from differences in subject matters.

Philosophical interpretation of science would, according to some, either be mere speculation, reflective of our lack of knowledge, or a matter of subjectivity and personal taste: therefore irrelevant for science. In a more permissive vein along the same line of reasoning, one might concede that there are interpretational issues and matters of debate in science, but maintain that they only concern the human aspects of science, the use humans make of science: matters of ethics or subjective meaning of concepts; interpretations, being quite independent of the truth itself that science discovers, do not or should not have any significant bearing on science. Debates would result from uncertainty and lack of knowledge, rather than being part of science.

## 3. Biting the bullet? Characterising 'philosophy'

Maybe these criticisms are not so off the mark. Maybe we should as philosophers just bite the bullet, and accept that we have managed to make ourselves irrelevant by disengaging from the latest concerns of science (a)—perhaps because we are not interested in science (c), or because we are not good at it (b). Of course, this would be an oversimplification. For there are plenty of philosophers who are interested in science, as well as scientifically well-informed. But for the next two paragraphs, I wish to entertain the thesis that maybe those criticisms are right, before I say a bit more about how I will use the word 'philosophy' in the rest of the paper. Scientists often lament that philosophers are ignorant of science, that they do armchair science, that they never test their theoretical conjectures, that philosophers make empirical claims that are known to be false, etc.

This may have been historically true of some philosophers like for example Hegel, and perhaps it would not be hard to find current books, written by philosophers, illustrating these shortcomings.[3] So yes, maybe Hawking's criticism—that one mischievous and catchy sentence, "philosophy is dead… Scientists have become the bearers of the torch of discovery in our quest for knowledge"—even if outrageously oversimplifying, does have bite.

But biting or not: it remains a false oversimplification. The rest of my essay will be concerned with how science needs philosophy. That should of course not make us forget the other fact—that philosophy needs science just as badly—but this essay will put that question aside.

At this point I should say a bit more about what I mean by 'philosophy'. Defining philosophy is not an easy task; and the nature of philosophy, or of the philosophical life, has been one of the traditional philosophical questions—it was in Athens, at any rate: but, fortunately, I will not be concerned with the nature of philosophy, as such, in this essay, but rather with parts of philosophy that are close to science.

I will be especially concerned with philosophy of science, more specifically with what one could call 'the philosophy of X', where X is a scientific discipline such as physics, chemistry or biology. General philosophy of science is of course particularly relevant for science, since it reflects on the nature and structure of scientific theories, and on the scientific process itself. But my main argument will be about the sub-disciplines of philosophy of science concerned with specific disciplines. In this way, philosophy of science fulfils various roles. It engages critically, at various levels, with the foundations, methods, and results of the sciences. Thus it not only makes explicit what is often only mentioned implicitly by scientific theories, but it also analyses the concepts and methods of scientific theories, and engages with the interpretation of their results. Furthermore, philosophy is, as we will see in section 4, sometimes used more constructively to develop new scientific theories—what I will call science's 'heuristic function'.

But saying that philosophy of science is relevant to science does not mean that the importance of philosophy is *limited* to philosophy of science. For philosophy of science itself builds on discussions in other parts of philosophy—not only the history of philosophy, but also ethics, epistemology, and metaphysics, to name a few. Thus, although my main argument is geared towards the relevance of philosophy of science, one should not lose these broader aspects of philosophy of science of sight. Subdivisions within philosophy are drawn for practical reasons, but when analysing specific problems they can also be artificial—as we will see in section 4b), where philosophy of physics requires discussing questions that belong to epistemology and ontology. Thus metaphysics, once banned by the logical positivists in the heyday of their youthful excesses, now thrives happily in analytic philosophy in ways that would have made Carnap, and even Quine, frown. But never mind the old glories—philosophy will never obey all your commands and prohibitions, and it will use whatever tools it can get hold of.

## 4. Science needs philosophy

I now get back to the response to the anti-philosophy arguments given in Section 2. What can one answer to these arguments, which seem to echo our most endearing notions and intuitions about the nature of science? Can we really deny that science and philosophy are two different worlds; that their subject matters and methods differ? Can we deny that science seeks to explain brute matters that are quite independent of human life? Can we deny the fact that unquestioned philosophical preconceptions have at times been hampering factors of scientific progress? Of course, we can't, as I hinted at in the previous Section: philosophy corrupts the youth—I think we all need to start from that. But that's only one side of the story, and not the most interesting part for us.

[3] I thank an anonymous referee for suggesting some of the above criticisms of current philosophy. For some examples, see Ladyman and Ross (2007: pp. 17-24).

As I will argue, the doctrine that philosophy is useless for science is not only false: it is also harmful for education, society, and ultimately science itself. I will do this by advancing three arguments for the usefulness of philosophy for the natural sciences. These arguments include refutations of the misconceptions presented in section 2. They are neither wholly original nor exhaustive, but they should be a first step towards the development of a synergetic relationship between philosophy and the natural sciences.

Given the tensions between science and philosophy, vividly expressed by physicists such as Stephen Hawking and Lawrence Krauss in recent works, trying to gain some clarity in this confused subject is by itself an important and urgent task.

**a) Why philosophy is useful (Ad 2a))**

**i. The fallacy of anti-philosophicalism**

Let me start with a simple contention that responds to a small, logical, part of the previous arguments: what I have called the fallacy of anti-philosophicalism and its refutation. The refuting argument boils down to something like this: "In order to argue that one does not need philosophy, one must do philosophy." Indeed, a convincing argument to the effect that "philosophy is useless for science" will necessarily entail the act of philosophizing. Even if 'useful' is a practical notion, arguing for the uselessness of discipline A for discipline B requires philosophical knowledge about B: one needs to argue that A is irrelevant to the subject matter, method, and goals of B. To declare categorically the uselessness of philosophy for science is therefore to have complete knowledge of the goals, method, and subject matter of science. But one can only argue about what those goals and subject matter should be by doing philosophy—more specifically, philosophy of science. Furthermore, we can only infer general statements about the usefulness of philosophy for science, from the study of a limited number of historical cases, by appending that study with a philosophical argument: hence by doing philosophy, in the way that historians and philosophers of science do it.

Does this debunk the argument about the death of philosophy? I submit that it does. For he who wants to insist on philosophy being useless for science must not try to rationally argue for this conviction, but must keep it as a matter of private opinion: for as soon as he starts to rationalize his view, he must start philosophizing. If there is some truth in that, as Hawking announces, philosophy is dead—and there may be some sense in which this is true—and that "Scientists have become the bearers of the torch of discovery in our quest for knowledge", then scientists can only do so by becoming philosophers of science themselves, hence resurrecting philosophy. Hawking acknowledges this by engaging in the discipline he has declared to be 'dead', thereby becoming a philosopher. Indeed, training in philosophy has at least this use, that it prevents us from being *bad* philosophers.

But, when arguing for the usefulness of philosophy, a logical argument is not necessarily the most convincing one. For it might lead us to replace the fallacy by a more cautious statement: “philosophy is useless for science, except for one thing: to argue that philosophy has no other use for science whatever.” Nevertheless, the fact that the former statement was false and the latter sounds arbitrary and contrived, leads us to question the soundness of an approach that declares philosophy to be close to useless. It might lead us to the idea that perhaps there is some genuine value in philosophy which is useful or even necessary for science and for scientists after all. I will defend the view that philosophy is useful to scientists, and that some amount of philosophical activity is necessary in order to construct a theoretical framework for doing science.

### ii. Paradigms and presuppositions (why science can’t live without philosophy)

The necessity of philosophy for science can easily be understood from a Kuhnian perspective on how science develops. Thomas Kuhn explicates progress in science not as a linear process of theoretical formulation and experimental verification or refutation of scientific theories, but in terms of revolutions and changes of paradigm (Kuhn 1962). A paradigm is for Kuhn not a cookbook recipe about the mathematical laws and mechanical workings of the universe or a set of equations and technical terms and procedures. Paradigms include ways of looking at the world, practices of instrumentation, traditions of research, shared values and beliefs about which questions are considered to be scientific. Nowadays we might want to stretch this concept even further to include institutional conditions, governmental constraints and market stimuli that may be supportive of particular paradigms[4]. Scientists working in different paradigms view the world in different ways, Kuhn has emphasized. Their basic assumptions about the kinds of entities there are in the world differ, as do the kinds of primary properties they attribute to those entities. Scientists working in different paradigms may disagree, as did Einstein and Bohr, about what makes a good theory or a good explanation; or about what it means to understand a problem. In other words, there are a wide range of ontological, epistemic, and ethical presuppositions weaved into any given scientific paradigm (for some examples of this, see Section 4b). If it is the case that a paradigm cannot come to birth, gain support, defeat its competitors, consolidate, and eventually die without such a set of explicit or at least tacit presuppositions, then presuppositions must be an intrinsic and necessary part of science regarded as a pursuit of truth. Such philosophical presuppositions are contributory to scientific theories, even if the theories are formally independent of them, because axioms cannot even be formulated without an agreement, taken from common and technical language, and justified within a wider paradigm, over what the terms mean and what kinds of entities they apply to; without implicit or explicit assumptions about how the terms relate to experimentally measurable quantities; without prescriptions for how the results of the theory can be verified or falsified. Paradigms also suggest meaningful goals and open questions for the theory. Thus philosophy plays a heuristic role in the discovery of new scientific theories (de Regt 2004): paradigms can function as guides towards the formulation of theories that describe entities of one type or another. As de Regt has cogently argued (see also the examples in the next Section), many great scientific innovators have at some point studied the works of philosophers and developed philosophical views of their own. This did not always happen very systematically, but the interest in philosophy developed by these scientists was at least above average and in turn had an important *heuristic function* in the formulation of new scientific theories (de Regt 2004).

[4] For the importance of tools and instrumentation, contexts, and power in different science cultures, see Galison 1996, 1997.

Implicit in the heuristic role of philosophy is also an important *analytic function,* as I stressed in Section 4.[5] One task of philosophy is to scrutinize the concepts and presuppositions of scientific theories, to analyse and lay bare what is implicit in a particular scientific paradigm. It is a philosophical task—one which is often carried out by physicists—to clarify the concepts of space, time, matter, energy, information, causality, etc. that figure in a given theory. This analysis is philosophical in so far as it makes explicit the implicit assumptions in the uses of these concepts: assumptions that scientific theories do not themselves normally state. Hence it moves beyond the point where the concepts appear as irreducible elements in the postulates of a theory. This analytic function should ultimately allow for a further step of integration, where the concepts of one science are related to the concepts of another.

The analytic function of philosophy might not only feed back into science, but become a starting point for philosophy itself: discovering what entities science assumes there to be in the world can be a useful starting point for philosophical reflection on nature. It seems key that philosophical stances on nature and science be compatible with the kinds of objects and relations that science finds. In the example given earlier: mechanistic philosophy did not admit the concept of action at a distance because the only forces allowed by the dominant philosophical paradigm were mechanical, hence the opposition to Newton's gravity theory; whereas Kepler's Pythagoreanism did allow for such a concept[6].

To summarize, then, some of the tasks for philosophy that we have found in relation to science:

1. To allow for, indeed to naturally incorporate into its own framework and build upon, the kinds of entities that science encounters in the world, and their properties and relations;[7]
2. To scrutinize the terms and presuppositions of science, i.e. to make explicit the implicit assumptions of scientific theories: to critically analyse and clarify what the terms used by science mean, how they are articulated, and what assumptions they require, as well as how they relate to the entities that philosophy argues there to be in the world;
3. To discover standards for what good theories, valid modes of explanation, and appropriate scientific methods are: to offer an epistemology that does not thwart, but stimulates scientific progress;
4. To provide ethical guidance and discover (broad) goals for science;
5. To point out and articulate the interrelations between concepts that are found in different domains of the natural sciences as well as the social sciences and the humanities;
6. To explain how observations fit in the broader picture of the world, and to create a language where scientific results and broader human experience can complement and mutually enrich each other.

[5] This 'analytic' function of philosophy does not strictly correlate with analytic philosophy. Both the analytic and continental traditions have of course been concerned with analysis of science and of its results.

[6] This holds true despite the fact that Kepler was one of the initiators of mechanistic science, and that also Newton in various ways held mechanical views. He regarded his theory of gravity as a phenomenological, inductively generalized law of nature that would nevertheless require further explanation as to its causes.

[7] See for examples the debates about the status of the wave-function in quantum mechanics: it is an important question whether the wave-function is a real entity existing in the world, or whether it merely represents the information about a system. This is a question that the formalism by itself does not answer, but nevertheless is important for how quantum mechanics is interpreted and used.

This list is neither exhaustive nor unique. Some of these general ideas will be instantiated in the two examples given in the next section.

The above points to a necessary relationship between science and philosophy. Science needs philosophy, as we have seen, in its two functions: heuristic, and analytic. Especially during changes of paradigm, philosophical debate will be part of the activities of science. None of this is to say that scientists need to be philosophers: most of them are not. So, philosophers may be drawn in at that point. But it is also not to say that professional philosophers should be doing all of the above tasks. Part of those tasks—surely 1 to 4—are often performed by scientists. Thus what I envisage here is a collaboration between scientists and philosophers. Indeed, I think we should be careful in distinguishing the disciplinary differences from the professional or individual ones. Saying that science, as a systematic theoretical and experimental study of the natural world, needs philosophy—which I have defined, in the analytic tradition, as the study of all the results of the sciences and humanities using the method of conceptual analysis—is not to say that each scientist requires philosophy. Philosophy may be merely a useful tool for scientists.

**b) Why the historical argument fails: quantum information, alive and kicking (Ad 2b))**

In this section, I give two examples where philosophical discussion has been genuinely contributory to science, along the lines discussed in 4a)ii. Before I do that, I will address the negative examples given in 2b)—examples where philosophy's influence was rather hampering: the iron clad of mechanistic philosophy and Plato's dictum that celestial motions should be along circles.

Working from 4a)ii we can now easily see that these examples in fact become a case in point: they illustrate the importance of philosophy for science. They make clear the need for having the right philosophical framework when doing science. If a conceptual enterprise such as philosophy were completely neutral, or indeed useless, to science, it could not be harmful to it in any important way either. But the fact is that: (A) some philosophical doctrines have been harmful for science while others have been productive; (B) it is impossible to have no philosophy at all (as I argued in 4a)); (C) the reason philosophy was harmful in some cases is because it was used in a positive way, according to its heuristic function from 4a)ii. And this heuristic function can indeed also be used positively. The correct course of action, then, is not to neglect philosophy—because, as per (A) and (B), philosophy can't be neglected—but to embrace its presence and to use it in an intelligent and positive way, as in (C). It follows from (A) to (C) that philosophy must be relevant to science in its own specific way, even if it is only in the manner of setting necessary intellectual preconditions of freedom of mind, of trust in the power of reason and of experimental observation, etc. History shows that it is hard for scientists to free themselves from outdated philosophical modes of thought. This highlights the importance of investing in having a philosophical framework that allows for the kinds of entities that science encounters in the world. Specific tasks for philosophy are as listed in 4a)ii.

Next we will study positive historical examples where 4a)ii is at work, thereby refuting the historical argument formulated in 2b). To refute the historical argument, it suffices to show one example where philosophy has been genuinely contributory to the progress of science. The example will be interesting in so far as it also sheds light on why it was that philosophy contributed to science, thus instantiating elements of 4a)ii. There are many such examples. Kepler[8] and Sommerfeld were both inspired by Pythagorean philosophical ideas when working out their models of the harmonies of the solar system and of the atom, respectively. Let me here concentrate on another, more recent, example. It concerns the current revolution in quantum information technology. In the past ten years we have seen the first commercialization of quantum randomness: the first bank transaction built on the basis of a code encrypted not by the usual algorithms of classical cryptography (which rely on unproven mathematical assumptions such as the difficulty in factorizing large prime numbers), but based on the new field of quantum cryptography: a technique for encoding messages based on the notion of entanglement between particles at long distances. Quantum cryptography has been successfully developed and commercialized by several groups over the past twenty years or so.

As it turns out, the quantum information revolution is rooted in the efforts of scientists who saw philosophical enquiry as a necessary step in their quest for knowledge. There are two key moments in the history of quantum mechanics when physical progress crucially depended on asking the right philosophical questions. Let me take these two episodes as case studies of the question how philosophical ideas influence science, in terms of philosophy's heuristic and analytic functions.

### i. Einstein versus quantum mechanics

In 1927, conflicting views on quantum physics started to crystallize. Towards the end of the 5th Solvay conference in Brussels, Werner Heisenberg declared quantum mechanics to be a "closed theory, whose fundamental physical and mathematical assumptions are no longer susceptible of any modification" (Bacciagaluppi 2006, p. 437). In doing so, Heisenberg was voicing the shared feelings of his colleagues Niels Bohr, Wolfgang Pauli, and Paul Dirac, also present at the conference. But Einstein and Schrödinger would have none of it: the Copenhagen interpretation—as the new view of quantum mechanics came to be known—had philosophical implications that they deemed undesirable. Among those properties was the lack of determinacy in physical quantities and events. Also, Heisenberg and co. seemed to introduce a possible role for human observers in the definition of the concepts that went into science.

A few years later, in 1935, Einstein, Podolsky, and Rosen made the nature of their discomfort with quantum theory explicit in a famous article that came to be known as the EPR thought experiment. They considered pairs of correlated particles separated at long distances. The possibility to measure a property (for example, the momentum) of the first particle automatically gives information about the value of that property for the second particle, without measuring that property for the second particle, since the particles are in a state of correlation. And the possibility to measure the complementary property (for example, the position) of the first particle would as well determine the value of that quantity for the second particle. But because of the assumption that measurements done on the first particle cannot affect the properties of the second particle (after all, the particles are well-separated), the second particle must have had the values of its position and momentum determined before any measurements were done on the first particle (since, according to the formalism of quantum mechanics, a measurement of the first particle determines the value of that property for the other particle, in both cases). Since, according to standard quantum mechanics, a particle cannot simultaneously have determinate values for both its position and its momentum, this means that quantum mechanics is an incomplete theory: for it does not predict properties for the second particle that, according to the argument, it can clearly have.

[8] For a visualisation of Kepler's model of the universe, see Katherine Brading's Digital Visualization Project: https://katherinebrading.wordpress.com/news/digital-visualization-project.

The EPR argument is philosophical in the sense explained earlier, in Section 3: for it analyses the foundations of quantum mechanics, trying to think clearly about the assumptions being made by standard quantum mechanics. But it also contains two substantive ontological assumptions. The first is what EPR call the 'criterion of reality' that if, using the formalism of quantum mechanics, one can predict with probability one the result of a measurement, then there is an element of physical reality corresponding to the physical quantity, with value equal to the predicted value of the measurement. The second assumption is what they call 'locality': namely, that elements of physical reality pertaining to one system cannot be affected by measurements performed on another system that is spacelike separated from the first.

Thus EPR's quest was both physical and philosophical. In addition to these two ontological assumptions, they also impose 'completeness' as an epistemic desideratum that a theory should satisfy: namely, that 'every element of the physical reality must have a counterpart in the physical theory'.

This led EPR to push the physical arguments farther than anybody had ever done before. The study of paradoxes borne out by thought experiments such as EPR has always played a major role in physics; but the resolution of such paradoxical situations almost invariably requires a philosophical stance about the principles and methods that are valued and deemed legitimate.

The EPR paper was truly philosophical in so far at it analysed and questioned the conceptual foundations of quantum theory. Especially EPR's construal of the notion of completeness, and their criterion of reality, are explicit epistemic and ontological positions.

Does this mean that Einstein was being professional philosophers while he worked on that paper? Of course not. One should distinguish *doing philosophy*—something that, like I said before, can be done by both physicists and philosophers—from one's professional label. Einstein was doing the philosophy that physics required at that point in time—and it was philosophy because he was reflecting on, and critically and constructively engaging with, the conceptual foundations of quantum theory. To do that, he needed philosophical tools. But he was of course also doing physics. So, by bringing philosophical methods into physics, he was advancing physics. I believe it is artificial, at such interdisciplinary intersections, to attempt to make too fixed a demarcation between physics and philosophy. Einstein was simply doing ground-breaking work that required methods from both fields.

### ii. Physics and the hippies

The next episode in this story of physics and philosophy took place many years later. After the publication of EPR, physicists continued to philosophize about the interpretation of quantum mechanics, but eventually the discussion died out. During the cold war, science and in particular physics gained much prestige. As class sizes grew, increasingly less time was spent on big questions and philosophical debates in the classrooms. While part of the reason for this decrease of attention on philosophical issues may have been pragmatic—philosophical discussions with large groups of students are hard to manage, and grading essay questions in exams is significantly more time consuming than computational questions—a vision was certainly at play about what education in science and technology should prepare students for. The interpretation of quantum mechanics was unlikely to prepare students who could provide societies with new gadgets or governments with new powerful weapons, whereas technical mastery of the formulas actually might. The old generation of physicists had received thorough training in the humanities—Werner Heisenberg once said "My mind was formed by studying philosophy, Plato and that sort of thing" (Buckley 1996, p. 6) and they had indulged in philosophical musings about the meaning of it all. Now the new generation of strong-headed physicists uttered the war whoop "Shut up and calculate" and instructed their students to rally behind their utilitarian flag. Making gadgets was the new goal of physics.

The instrumentalist view of science regnant during the decades after the war is explained by Lee Smolin as follows: "When I learned physics in the 1970s, it was almost as if we were being taught to look down on people who thought about foundational problems. When we asked about the foundational issues in quantum theory, we were told that no one fully understood them but that concern with them was no longer part of science. The job was to take quantum mechanics as given and apply it to new problems. The spirit was pragmatic; "Shut up and calculate" was the mantra. People who couldn't let go of their misgivings over the meaning of quantum theory were regarded as losers who couldn't do the work." (Smolin 2007, p. 312).

But instrumentalism had to give way to other kinds of motivation for doing physics. Economic recession, budget cuts, and the decrease in the number of physics jobs made class sizes decrease again. Physicists once again had the time to think about the meaning of what they were doing. In a second, seemingly unrelated line of developments, the CIA, afraid that Americans would lag behind the Soviets, decided to fund laser physicist Harald Puthoff at Stanford University's SRI lab in Menlo Park, California, for the study of psychic phenomena. Additional money came from NASA. Soon Puthoff would be associated with a third strand of events around the Bay Area. A dubious consortium of hippie physicists and quasi-crackpots formed an unlikely discussion group. They alternated their musings about all things quantum and the meaning of life with drinking parties and psychedelic drug use. They came to be known as the Fundamental Fysiks Group and eventually found a generous patron in self-help industry forerunner and multi-millionaire guru Werner Erhard. One goal of the hippie scientists was to use quantum mechanics for superluminal (faster-than-light) communication. This would include communication with their deceased colleagues. Needless to say, many of their arguments were misguided, but their contribution to physics was of lasting endurance. They not only put the interpretation of quantum mechanics on the research and teaching agenda; they analysed the EPR arguments and the important contributions to this discussion made by John Bell, David Bohm, and others, which had escaped the attention of scientists until then; they helped clarify the issues at stake, developed new thought experiments of their own, and raised awareness that quantum nonlocality might be useful in long-distance communication. Save the crucial (wrong) conclusion that superluminal communication was possible, several set-ups and techniques the hippies considered did not differ significantly from the ones that quantum communication uses nowadays. As David Kaiser has argued (Kaiser 2012, p. xxiii), "The group's efforts helped to bring sustained attention to the interpretation of quantum mechanics back into the classroom. And in a few critical instances, their work instigated major breakthroughs that—with hindsight—we may now recognize as laying crucial groundwork for quantum information science."

Like Einstein, the Fundamental Fysiks Group worked at the intersection of physics and philosophy. They brought philosophical methods and literature to bear on problems in physics, and as such they did the kind of work that I argue physics periodically needs—regardless of who does that work, whether it is the physicists themselves or the professional philosophers.

The two examples illustrate some of the tasks of philosophy for science listed in section 4a)ii. Progress in fundamental issues such as entanglement and quantum communication stemmed from physicists' willingness to engage in debates about ontological and epistemic issues such as the role of the observer, the completeness of the mathematical description of nature, the desiderata for a good description of nature, and so on. Progress not driven by such philosophical questions is hard to imagine in this case; the philosophical debate that actually took place acted as a positive, guiding force that pushed science further; fuelled by the posing of legitimate and relevant philosophical questions in their quest for new physics, by their being insistent on philosophical clarity and coherence rather than content with just technical mastery of the formulas, which was the trend of the day.

**c) Synergy between science and philosophy (on objects and methods) (Ad 2c))**

There are two sides to the objection regarding the difference between science and philosophy as forms of scholarship: subject matters on the one hand, methodology on the other.

I will be brief about the distinction in subject matter. Philosophy studies every subject matter that the sciences also study (recall my 'philosophy of X' from section 3), but it does this with different aims and methods. The universe, possible universes other than our own, elementary particles, life, are all subjects of concern for both natural science and philosophy. Therefore, on those overlaps science and philosophy cannot be distinguished on the basis of their subject matters alone. The difference is often sought in their formal objects and methodologies: the earlier mentioned distinction between *erklären* and *verstehen* could be reframed as the statement that the natural sciences seek explanations in the modes of causal efficacy and material causation, whereas philosophy is interested in formal analysis, goals, and intentionality. This difference in methodology is often summed up by the mantra: 'philosophy asks why-questions, science asks how-questions'. And I agree: their methods are different, and their aims (in particular, the specific interest from which they study 'the same subject matter') are also different. But I also submit that any such division cannot be made once and for all—the division is both vague at any point in time, as well as dynamical.

By declaring that there is such a division of intellectual activities, natural scientists and philosophers can comfortably go about their work without competing or stepping on each other's shoes. But, as I am suggesting, the mantra is as comfortable as it is lacking in accuracy in fully reflecting the nature of the relationship between science and philosophy. Agreed: science and philosophy are in principle different forms of scholarship. For established fields of science such as classical mechanics or electromagnetism, there may be much truth in the statement that science is practically interested in how-questions, defined by the framework of the particular paradigm one is working in. But that is so only because a number of why-questions have been answered within the wider paradigm and are not being questioned any further. When paradigms are in the making, there is no clear-cut distinction between the scholar asking the how-questions and the scholar asking the why-questions. Any how-question may lead us to a why-question, and any answer to a why-question may lead us to answers to multiple how-questions. When placed in front of a why-question in the quest for a new theory, the scientist cannot retreat into the shell of specialism. He or she must struggle with the question using whichever intellectual means are available. He or she may need to establish, as the founding fathers of quantum mechanics attempted to do, what a measurement is before they can convincingly argue that there is such a thing as uncertainty in the microscopic world. The scientific quest presupposes having a number of philosophical issues settled first: or, at least, it presupposes engaging with the various conceptual options, and taking a stance on them. In so doing, the subject matters and methods of philosophers and of scientists become entangled: the relationship between science and philosophy becomes dynamical.

This is particularly true in our time, when science has expanded into realms—from far-away galaxies to the multiverse to neuroscience to molecular engineering—that were unknown territory just a number of decades before. Science is aimed at truth about the natural world, and although methodological distinctions can be made formally, one must be aware of their limitations: in particular, it would be wrong to conclude that a methodological distinction allows us to dismiss philosophy for the sake of science.

This brings us to another point: if science needs philosophy, scientific results should also be the starting point of philosophical reflection about nature. It is probably here that Hawking's criticism of philosophy has an important core of truth to it (see footnote 1).

There is another reason why science needs philosophy. Scientific knowledge is not technical specialism cut off from the rest of human knowledge. The moment this happens would signal the forthcoming death of science. Scientific results constitute knowledge to be integrated into the broader human quest for answers about ourselves and about the universe. Philosophy helps the scientist articulate her findings in a kind of knowledge that can be shared with others, not experts in her field; it will help her discuss with other intellectuals and contribute to the general human task of getting to know the world and ourselves better.

To summarize my main argument so far: the relationship between science and philosophy may be in bad shape, and philosophy may be in bad shape, but it cannot be dead as long as we are trying to understand the universe around us. Historically, philosophy has been very influential for science, as has science been for philosophy. Any instances where philosophy had a negative effect on science in fact contribute to highlighting the importance of thinking carefully about the relation between philosophy and science. Science cannot do without philosophy because there are philosophical stances implicit in the presuppositions and goals of any scientific paradigm and in how theories are connected to reality: and it is the task of philosophy of science to critically engage with those presuppositions. Thus science needs philosophy to scrutinize those presuppositions, stances, and goals. And philosophical tools are sometimes required to make progress—as the EPR and quantum information revolution illustrate. Finally, science requires philosophy to connect its findings to the rest of human knowledge. Philosophy can act as a language connecting disciplines that are far away from each other.

Since the subject matters of science and of philosophy are partially overlapping, formal or methodological distinctions between science and philosophy only have limited ranges of applicability and certainly do not imply independence of the two disciplines. In other words, the boundaries between science and philosophy are not water-tight, nor should they be.

## 5. Liberal arts and sciences: freeing the mind

Having argued, at the end of the previous section, that science as such needs philosophy, I will now look at the implications of this statement for education. That is, I would like to add a few reflections about how scient*ists* need philosophy, and how this is to be reflected in education.

Let me start by examining what does and does not follow from what we have established so far. From the assertion that science needs philosophy in some way it does not follow that each individual scientist should be a skilled philosopher, or in fact should have any kind of developed skill in philosophy. A scientist faced with a philosophical question in the course of her research might choose to neglect it and still do a relatively good job at her research, at least for some time. Also, despite the fact that every scientist has a philosophy that is at least weaved into the presuppositions and goals of the given theory or paradigm that the scientist works in, perhaps appended with her own private reflections, it is true that science can be done for the sake of science with neglect of the philosophical presuppositions and for exclusively utilitarian goals. Obviously, utilitarian values do not offer a sustainable basis for science as a whole and for maintaining public trust in the meaningfulness of fundamental research. But for the individual scientist, they might just suffice. Furthermore, even in the case that the scientist has her own philosophical views, she is free to keep them private and not let them interfere with the research she is doing. In fact, scientists may work together on the same scientific problem while sustaining different ontological or epistemic presuppositions. Philosophy may be even less relevant for the applied scientist (although, especially for her, ethical issues will be important!). So, for all practical purposes, the individual scientist might get away with neglecting philosophy. What use, one might cynically enquire, will the laser physicist have in formal training in philosophy? Even taking the point that every scientist in fact makes use of philosophical thought of one kind or another—a set of ideas about the scientific practice, about the nature of the objects and relations that constitute her subject matter, etc.—one may still argue that it is enough for the individual scientist to work within the philosophical framework of a specific paradigm; to employ, in her daily work as a scientist, the intuitions that she internalized in the context of the specific paradigm or tradition in which she was trained. There is no need for receiving specific training in philosophical matters. Thus philosophy courses of the kind I have in mind cannot be seen as necessary prerequisites for any single scientist. But I argue that they are *useful* for them, and that scientists would benefit from them: and so, that science programmes *ought to have such courses*—again, without going into details, which would require a separate paper.

So, this suggests the following question. Shouldn't the education of future scientists somehow reflect the connection that we have found between the sciences and philosophy? Indeed, particularly in the context of liberal arts and sciences, it is key that education reflects that connection. Science students in modern liberal arts and sciences programs should receive training in philosophy specific to their particular sciences. The kind of training I am arguing for here goes beyond general courses such as logic and philosophy of science, which are very important and are already part of some liberal arts and sciences curricula, as electives at least. It also goes beyond ethics, which is obviously an important training for scientists—although here one should go beyond the theoretical cocktail-party way in which some of these courses are taught, since their relevance often escapes the student. Perhaps such courses should be based more on actual scientific episodes and practices. But ethics is in itself a very large subject, and I have something else in mind here that more directly relates to my case studies: namely, philosophical reflection specific to each of the sciences, in fact specific to each particular science course a student takes. I mean courses such as 'The Philosophy of X', where X is a discipline or a collection of related disciplines (see the discussion in section 3). And I would argue that such materials could also be part of every science course, rather than separate courses, and so are best taught by scientists. If one is intrepid, one might wish to add a course on theory construction: but I admit, this will not be easy, though it could be very beneficial at the graduate level.

Historically, it has been a goal of liberal arts and sciences education to educate the social, political, and intellectual elites. In our century, the liberal arts and sciences are often advertised in somewhat different, but related terms: 'training the leaders of the future who can solve global problems' is something one often hears as part of the institutional rhetoric about liberal arts and sciences. Selective admission procedures, small class settings, and emphasis on basic logical, argumentative, and rhetorical skills do confirm this vision. Clearly, some of these leaders will also be leaders in their respective scientific fields, whether in applied or in fundamental science. So, if the liberal arts and sciences aim at training the intellectual elites of the future, in particular they should be interested in the scientists who can really make a difference in research and scientists who will be the leaders of other scientists. More precisely; I will take a useful practical distinction made by Lee Smolin, even if I don't agree with the broad-brush way Smolin applies it to string theory, nor with the details of his comparison with Kuhn's idea of revolutionary science. The distinction goes back to Einstein, who wrote in a letter (letter to Robert A. Thornton 7 December 1944, EA pp. 61-574): "I fully agree with you about the significance and educational value of methodology as well as history and philosophy of science. So many people today - and even professional scientists - seem to me like someone who has seen thousands of trees but has never seen a forest. A knowledge of the historical and philosophical background gives that kind of independence from prejudices of his generation from which most scientists are suffering. This independence created by philosophical insight is - in my opinion - the mark of distinction between a mere artisan or specialist and a real seeker after truth."[9] In this rich quotation, Einstein argues, as he did in many other occasions, for the significance of training in the history and philosophy of science, which gives the scientist independence of thought, which is precisely the kind of liberation of the mind that liberal arts programs also seek. Second, he calls this freedom of mind the mark of distinction between a mere specialist and a real seeker after truth. Smolin explicates this as follows. He divides scientists into seers and scientists who are master craftspeople. The seers are the ones leading the way, the ones who can see the whole forest, in Einstein's words 'the seekers after truth'. The master craftspeople are the ones who are very good at their particular trade, but have never seen a forest—be out of lack of interest or lack of sight. Smolin relates these two categories of scientists to the two types of science in Kuhn—normal science and revolutionary science.[10] In normal science, all the details of a given paradigm are explored and worked out. This is mainly the master craftspeople's work. They explore the mine, excavate the tunnels, take out the valuable jewels in a mine that was found and planned by others. Revolutionary science, on the other hand, is the task of going into new territory, of doing the exploratory work required to establish radical new ideas; that is the work of the seers, the people who can think out of the existing paradigm—although never entirely—who can point out weaknesses in theories and propose new ways forward. Freedom of mind, among other things, is one of the characteristics of such scientists, and knowledge of history and philosophy contribute to that free way of thinking. If liberal arts and sciences programs advertise themselves as forming the leaders of the future, shouldn't they be seeking to form master craftspeople as well as seekers, searchers of truth? Shouldn't they be the breeding ground for scientists with a certain capacity of independence from prejudice and from the opinion of the majority as well as the ability to persuade others to pursue their radical ideals?

[99] Quoted by Smolin (2007), p. 310-311.

[10] I take Smolin's identification of the contrast of seers vs. master crafspeople with revolutionary vs. normal science to be merely a suggestive analogy. For there are important historical disanalogies too, which nevertheless do not militate against the point I am making about education.

## Acknowledgements

I thank the organizers of the conference *Rethinking Liberal Education,* where this paper was presented, for an inspiring conference. I also thank Dennis Dieks and Jeroen van Dongen for a long-term collaboration which has helped shape some of the ideas presented in this essay. I would also like to thank Palmyre Oomen and Rudi te Velde for discussions on these topics as well as thoughtful comments on the manuscript. I thank Jeremy Butterfield and two anonymous referees for their comments on the manuscript.